\documentclass{WileyChemistry-template}

\title{\raggedright Hybrid Functional and Plane Waves based {\textit {Ab Initio}} Molecular Dynamics Study of the Aqueous Fe$^{2+}$/Fe$^{3+}$ Redox Reaction}

\author{
\begin{minipage}{\textwidth}
	Sagarmoy Mandal,\textsuperscript{[a,b,c]} 
	Ritama Kar,\textsuperscript{[a]} 
	Bernd Meyer,\textsuperscript{[b,c]} 
	Nisanth N. Nair*\textsuperscript{[a]}
\end{minipage}
}

\newcommand{\affiliation}{
\begin{itemize}

\item[{[a]}] Dr. S. Mandal, R. Kar, Prof. Dr. N. N. Nair*\\
Department of Chemistry, Indian Institute of Technology Kanpur (IITK), 208016 Kanpur, India\\
E-mail: nnair@iitk.ac.in

\item[{[b]}] Dr. S. Mandal, Prof. Dr. B. Meyer\\
Interdisciplinary Center for Molecular Materials and Computer Chemistry
Center, Friedrich-Alexander-Universit{\"a}t Erlangen-N{\"u}rnberg (FAU),
N{\"a}gelsbachstr. 25, 91052 Erlangen, Germany\\

\item[{[c]}] Dr. S. Mandal, Prof. Dr. B. Meyer\\
Erlangen National High Performance Computing Center (NHR@FAU), Friedrich-Alexander-Universit{\"a}t Erlangen-N{\"u}rnberg, Martensstr. 1, 91058
Erlangen, Germany\\

\end{itemize}
}

\renewcommand{\dedication}{
	\begin{minipage}{\textwidth}
	
	\end{minipage}
}

\renewcommand{\abstract}{
Kohn-Sham density functional theory and plane wave basis set based {\textit {ab initio}} molecular dynamics (AIMD) simulation is a powerful tool for studying complex reactions in solutions, such as 
electron transfer (ET) reactions involving Fe$^{2+}$/Fe$^{3+}$ ions in water.
In most cases, such simulations are performed using density functionals at the level of Generalized Gradient Approximation (GGA).
The challenge in modelling ET reactions is the poor quality of GGA functionals in predicting properties of such open-shell systems due to the inevitable self-interaction error (SIE).
While hybrid functionals can minimize SIE, AIMD at that level of theory is typically 150 times slower than GGA for systems containing $\sim$100 atoms. 
Among several approaches reported to speed-up AIMD simulations with hybrid functionals, the noise-stabilized MD (NSMD) procedure, together with the use of localized orbitals to compute the required exchange integrals, is an attractive option. 
In this work, we demonstrate the application of the NSMD approach for studying the Fe$^{2+}$/Fe$^{3+}$ redox reaction in water. 
It is shown here that long AIMD trajectories at the level of hybrid density functionals can be obtained using this approach.
Redox properties of the aqueous Fe$^{2+}$/Fe$^{3+}$ system computed from these simulations are compared with the available experimental data for validation.
}

\newcommand{\keywords}{
    Electron transfer \textbullet\ 
	Hybrid functionals \textbullet\ 
	Marcus theory \textbullet\
	Redox reaction \textbullet\ 
	Reorganization free energy
}

\begin{document}
%%%%%%%%%%%%%%%%%%%%%%%%%%%%%%%%%%%%%%%%%%%%%%%%%%%%%%%%%%
%%%%%%%%%%%%%%%%%%%%%%%%%%%%%%%%%%%%%%%%%%%%%%%%%%%%%%%%%%
%%%%%%%%%%%%%%%%%%%%%%%%%%%%%%%%%%%%%%%%%%%%%%%%%%%%%%%%%%

\twocolumn[\vspace{-1.5cm}\maketitle\vspace{-1cm}
	\textit{\dedication}\vspace{0.4cm}]
\small{\begin{shaded}
		\noindent\abstract
	\end{shaded}
}

\newcommand{\gga}{{\bfseries BLYP} \xspace}
\newcommand{\hyb}{{\bfseries B3LYP} \xpsace}

\begin{figure} [!b]
\begin{minipage}[t]{\columnwidth}{\rule{\columnwidth}{1pt}\footnotesize{\textsf{\affiliation}}}\end{minipage}
\end{figure}

%%%%%%%%%%%%%%%%%%%%%%%%%%%%%%%%%%%%%%%%%%%%%%%%%%%%%%%%%%
%%%%%%%%%%%%%%%%%%%%%%%%%%%%%%%%%%%%%%%%%%%%%%%%%%%%%%%%%%
%%%%%%%%%%%%%%%%%%%%%%%%%%%%%%%%%%%%%%%%%%%%%%%%%%%%%%%%%%

%%%%%%%		 Main Text			%%%%%%% 

\section*{Introduction}
\label{introduction}

Electron transfer (ET) reactions are very important in many chemical and biological processes and are subjected to numerous experimental and theoretical investigations.\cite{book_et}
In this regard, aqueous ferrous-ferric ET has attracted many  theoretical\cite{ET_Fe_1,ET_Fe_2,ET_Fe_3,ET_Fe_4_Sit,ET_Fe_5,ET_Fe_6,ET_Fe_7,ET_Fe_8,ET_Fe_9,ET_Fe_10,ET_Fe_11,ET_Fe_12,ET_Fe_13,ET_Fe_14,ET_Fe_redox_potential,lambda_PSit,lambda_Fe_blumberger} studies as a prototype of more complex ET processes.
In water solvent, both ferrous and ferric ions remain in stable hexaaqua complex form and build a well defined solvent structure around them. 
Marcus has developed a very powerful theory\cite{marcus_1,marcus_2,marcus_3,marcus_4} for describing the rate of ET from the electron donor to the electron acceptor in solution.
%
%He found that the role of solvent polarization is very important in these kind of ET events 
%
%and used the vertical energy gap as the reaction coordinate to estimate the rate of ET reactions.
%
The role of solvent polarization is very important in these ET events 
and the vertical energy gap was used as the reaction coordinate to estimate the redox potential and the rate of ET reactions.
The reactant and product states involved in the ET process are described by two diabatic free energy surfaces.
%
%According to this theory, the surrounding solvent environment responds in a linear fashion with the amount of charge transfer.
%
%Due to this 
Using linear response theory, the free energy surfaces for the reactant and product states become parabolic functions of the vertical energy gap.
%
%An importance assumption of this theory is that 
%
Marcus theory has been successfully applied in many computational studies, and the linear response assumption was found to be valid.\cite{ET_review_blumberger_PCCP,ET_review_blumberger,ET_review_MARCUS_1,ET_review_MARCUS_2,ET_review_Sutin}
%
%With this theory, many ET systems were modelled and 
%These results match quite well the the experiments for many ET systems, which were modeled computationally with the help of this theory.

%Also, marcus theory helps us to 
For applying simulation methods to model ET reactions in solution,
Warshel and co-workers\cite{Warshel_1,Warshel_2,Warshel_3,Warshel_4} have devised a strategy to use force field based classical molecular dynamics (MD) simulations.
%to study these ET reactions in the framework of Marcus theory.
%
These statistical mechanics based theoretical results were fairly successful in predicting the experimentally observable redox properties.
However, for a better understanding of the molecular level details of the solute-solvent structure, Kohn-Sham density functional theory (KS-DFT) based {\em ab initio} MD (AIMD) simulations are required.   
KS-DFT can give an accurate description of the solvent structure around the solute, leading to a better estimate of the vertical energy gap.\cite{sprik_Ru_ET_qm_vs_class}
However, it is practically difficult to model the whole electron transfer process with KS-DFT due the involved technical difficulties.
In an attempt to avoid this problem, Sprik and co-workers\cite{sprik_Ag_ET_bomd,sprik_Ag_ET,sprik_Cu_Ag_ET,sprik_quinones_ET,sprik_Ru_ET_qm_vs_class,sprik_Ru_ET,sprik_TTF_TH_ET,sprik_tyrosine_tryptophan_ET} proposed to model only half reactions instead of the full reaction.
It has been shown that the Marcus theory equally holds for half reactions and the results match well with experiments.

A redox property of interest within the Marcus theory is the reorganization free energy ($\lambda$), which is defined as the free energy required to distort the ion from the most probable configuration of one state to the most probable configuration of the other state.
Previous computational studies on the aqueous ferrous-ferric ET reaction reported $\lambda$ as 3.6 eV\cite{ET_Fe_1} and 3.57 eV.\cite{ET_Fe_3} 
Experimental value of $\lambda$ is 2.1 eV,\cite{lambda_expt} and thus
these classical force field based calculations have overestimated the value of $\lambda$.
%{\color{red} which is found to be 2.1 eV.\cite{lambda_expt}}
%
One of the reasons for this discrepancy might be the lack of electronic polarization in the force field description of the solvent.\cite{ET_review_blumberger_PCCP}
%
%In these kinds of study, the solvent effects and finite temperature fluctuations are very important.
%
%
Thus, for a better quantitative result, KS-DFT based AIMD simulations are required, which can model the electronic polarizability of the solvent and the solute-solvent interaction more accurately.
However, until now, these systems were described mostly with Generalized Gradient Approximation (GGA) functional based DFT simulations.
Typically, GGA functional based AIMD simulations are inadequate in describing the electronic structure of open-shell systems, because they suffer from the self-interaction error (SIE).\cite{Chemist's_Guide,Science_DFT_limitations,PRB_SIC,PRL_DFT_errors}
Due to this error, these functionals tend to erroneously delocalize the excess electron over the donor and acceptor, and thus produce wrong estimates of vertical energy gaps, redox potentials and reorganization energies.\cite{ET_review_blumberger_PCCP,ET_Co_AIMD,Redox_chemistry_water,Electrochemistry_QMMM,redox_potentials_AIMD}
To address the problem of SIE, Marzari and co-workers formulated a penalty density-functional approach,\cite{ET_Fe_4_Sit} and applied it to study the aqueous ferrous-ferric ET reaction.
The value of $\lambda$ was found to be 2.0 eV in their study, which has excellent agreement with the experimental result.
On the other hand, Sit and co-workers\cite{lambda_PSit} have reported $\lambda$ as 2.28 eV from a constrained DFT study, and 
Blumberger and co-workers\cite{lambda_Fe_blumberger} have found $\lambda$ to be in the range of 1.70 - 1.95 eV from DFT/continuum calculations.
%of the 6- to the 18-water-molecule model for aqueous Fe$^{2+}$/Fe$^{3+}$ ions.

In our present work, we use hybrid functionals.\cite{Chemist's_Guide,Martin-book,JCP_B3LYP,JCP_PBE0,JCP_HSE}
%, where a certain percentage of Hartree-Fock (HF) exchange energy is mixed with the GGA exchange energy.
%
Hybrid functionals are known to reduce SIE and improve the quality of predicted redox properties.\cite{ET_Co_AIMD,Redox_chemistry_water,Electrochemistry_QMMM,redox_potentials_AIMD,cdft_cp2k_blumberger,Metallocene_IP,redox_Fe_S,CALVOCASTRO_2015,cdft_jcp_blumberger,ET_Co_IC,Ti-Cu_JPCB}
However, hybrid functional based simulations are at least two orders of magnitude slower than the GGA functional based calculations.\cite{JCP_HFX_Voth} 
Thus, AIMD simulations with hybrid functionals are rarely used\cite{cdft_cp2k_blumberger,Electrochemistry_QMMM} to study redox reactions.
Recently, we have proposed a few methods to speed up hybrid functional and plane waves (PWs) based AIMD simulations.\cite{JCP_sagar,JCP_2019_sagar,sagar_JCTC,sagar_JCC,sagar_JCC_2022}
These methods were successfully applied to model chemical reactions in explicit solvent. 
Here, we use one such method, namely the noise stabilized molecular dynamics (NSMD).\cite{JCP_sagar} %to study the current system under 
%NN considerations.
%consideration.
%
Our earlier implementation is modified to take care of spin polarised systems.

We model here the ferrous-ferric electron transfer process following the strategy of Sprik and co-workers.
KS-DFT based AIMD simulations were performed to compute the redox potential and the solvent reorganization free energy.
We benchmarked the performance of two different levels of density functionals, in particular, BLYP (GGA) and B3LYP (hybrid), for the prediction of redox properties.
Additionally, for a fair comparison with the experimental results, we estimated the error due to finite size effect. % of our system.  
%We performed both GGA and hybrid based simulations and compared their accuracy in terms of the predicting power.
%
%The dependence of the solvation structure of the ions on the level of functionals were also compared in detail.
%

%%%%%%%%%%%%%%%%%%%%%%%%%%%%%%%

\section*{Methods and Models}
\label{methods_and_models}
%==============================================================

\subsection*{Theory of Electron Transfer Reactions}
%The classical statistical mechanism formalism of Warshel can be followed to model 
%
%To model the aqueous ferrous-ferric redox reaction, we follow the methodology that was introduced by Sprik and co-workers\cite{sprik_Ag_ET_bomd,sprik_Ag_ET,sprik_Cu_Ag_ET,sprik_quinones_ET,sprik_Ru_ET_qm_vs_class,sprik_Ru_ET,sprik_TTF_TH_ET,sprik_tyrosine_tryptophan_ET} employing DFT based AIMD simulations.
%
%In this approach, instead of considering the complex full redox reaction, we focus on a half redox reaction.
%
%This method was successfully applied to study various redox half reactions,\cite{blumberger_Ru_ET_JACS,blumberger_Mn_ET_JPCB,sprik_M_ET_JPCB,sprik_Ru_ET_JCTC,Ensing_flavin_ET_JPCB,ensing_lumiflavin_ET_JCTC,ensing_Ru_ET_faraday_dis,rosso_ET_JPCA,parson_ET_JPCB,VANDEVONDELE_ET_AIMD,Tateyama_JCP_ET1,Tateyama_JCP_ET2} and it has been shown that the Marcus theory equally holds for this half reaction model.
%
%In this section, we briefly present the key equations that are useful for our analysis in the \nameref{results_discussion} section.
%
%
%
This work focuses on the
the following redox reaction 
\begin{equation}
\label{reaction}
{\text {Fe}}^{2+} (\text{aq}) \leftrightharpoons {\text {Fe}}^{3+} (\text{aq}) + {\text {e}}^-  \enspace .
\end{equation}
We follow the methodology that was introduced by Sprik and co-workers\cite{sprik_Ag_ET_bomd,sprik_Ag_ET,sprik_Cu_Ag_ET,sprik_quinones_ET,sprik_Ru_ET_qm_vs_class,sprik_Ru_ET,sprik_TTF_TH_ET,sprik_tyrosine_tryptophan_ET} employing DFT based AIMD simulations.
In this approach, a half-cell redox reaction is considered and has been applied to various problems.\cite{blumberger_Ru_ET_JACS,blumberger_Mn_ET_JPCB,sprik_M_ET_JPCB,sprik_Ru_ET_JCTC,Ensing_flavin_ET_JPCB,ensing_lumiflavin_ET_JCTC,ensing_Ru_ET_faraday_dis,rosso_ET_JPCA,parson_ET_JPCB,VANDEVONDELE_ET_AIMD,Tateyama_JCP_ET1,Tateyama_JCP_ET2} 
%
%This method was successfully applied to study various redox half reactions,\cite{blumberger_Ru_ET_JACS,blumberger_Mn_ET_JPCB,sprik_M_ET_JPCB,sprik_Ru_ET_JCTC,Ensing_flavin_ET_JPCB,ensing_lumiflavin_ET_JCTC,ensing_Ru_ET_faraday_dis,rosso_ET_JPCA,parson_ET_JPCB,VANDEVONDELE_ET_AIMD,Tateyama_JCP_ET1,Tateyama_JCP_ET2} and it has been shown that the Marcus theory equally holds for this half reaction model.
%where Fe$^{2+}$ and Fe$^{3+}$ denote 
%
%
%We model our system with a single ion (either Fe$^{2+}$ or Fe$^{3+}$) solvated with explicit water molecules.
%
%Thus, in \cref{reaction}, Fe$^{2+}$ and Fe$^{3+}$ refer to the whole system in reduced and oxidized form, respectively.
%
%Now, 
The central quantity of our interest is the vertical energy gap, $\Delta E$, defined as
\begin{equation}
\label{gap}
\Delta E ({\mathbf{R}}) = E_{{\text {Fe}}^{3+}} ({\mathbf{R}}) - E_{{\text {Fe}}^{2+}} ({\mathbf{R}}) \enspace ,
\end{equation}
where $E_{M} ({\mathbf{R}}), ~{M} = {\text {Fe}}^{2+}$ or ${\text {Fe}}^{3+}$, is the ground state potential energies of the system in reduced and oxidized states
%and
%including the solvent molecules.
%
for the given atomic configuration ${\mathbf{R}}$.
%is the ionic configuration of the system.
%
%
Within the framework of Marcus theory, $\Delta E$ 
can be used as the reaction coordinate to describe the free energy changes associated with the redox reaction.\cite{Warshel_1,Warshel_2,Tachiya_ET}
%
%
%Moreover, this theory assumes that the solvent environment responds in a linear fashion to the change in solute oxidation state.
%
%With this assumption, 
Further, the probability distribution of the vertical energy gap $\Delta E$, $P_{{M}} (\Delta E)$, is Gaussian shaped: 
\begin{equation}
    P_{{M}} (\Delta E) = \frac{1}{\sigma_{\Delta E}^{M}  \sqrt{2\pi}}\exp \left [ -\frac{(\Delta E - \left \langle \Delta E \right \rangle_{M})^2}{2 (\sigma_{\Delta E}^{M})^2} \right ] \enspace .
\end{equation}
Here, $\left \langle \Delta E \right \rangle_{M}$ is the average of $\Delta E$ %during a AIMD simulation of the system in 
and $\sigma_{\Delta E}^{M}$ is the standard deviation of the distribution, for the redox state $M$.
Now, the free energy surface for the system in state $M$ can be constructed as a function of the energy gap as 
\begin{equation}
\label{fes_con}
F_{M} (\Delta E) = -k_{\text B} T \ln P_{M} (\Delta E) + {\text {constant}} \enspace ,
\end{equation}
where $k_{\text B}$ and $T$ are the Boltzmann constant and the temperature, respectively.
The Gaussian shape of the probability distribution $ P_{{M}} (\Delta E)$ assures that the diabatic free energy plots for different oxidation states are parabolic, in accordance with Marcus theory.
%

%The reorganization free energy is defined as the energy required to distort the ion from the most probable configuration of one state to the most probable configuration of the other state while staying on the same free energy surface.
%
If $\Delta E_{M}^{\text {min}}$ is the value of $\Delta E$ for which $F_{M}(\Delta E)$ is minimum, then
%in the free energy curve of state $M$.
%
the reorganization free energy for the system in the reduced state is given by
%%can be written as 
\begin{equation}
\label{lambda2}
\lambda_{{\text {Fe}}^{2+}} = F_{{{\text {Fe}}^{2+}}} (\Delta E_{{\text {Fe}}^{3+}}^{\text {min}}) - F_{{{\text {Fe}}^{2+}}} (\Delta E_{{\text {Fe}}^{2+}}^{\text {min}})
\end{equation}
and for the oxidized state is 
\begin{equation}
\label{lambda3}
\lambda_{{\text {Fe}}^{3+}} = F_{{{\text {Fe}}^{3+}}} (\Delta E_{{\text {Fe}}^{2+}}^{\text {min}}) - F_{{{\text {Fe}}^{3+}}} (\Delta E_{{\text {Fe}}^{3+}}^{\text {min}}) \enspace .
\end{equation}
%
%%Interestingly, 
The choice of $\Delta E$ as the reaction coordinate leads to the following equation\cite{Warshel_1,Tachiya_ET}
\begin{equation}
\label{lineqn}
   F_{{{\text {Fe}}^{3+}}}(\Delta E) - F_{{{\text {Fe}}^{2+}}}(\Delta E) = \Delta E  \enspace , 
\end{equation}
which relates the free energy surfaces of both states.
%
%Therefore, 
The free energy surface for the system in state Fe$^{3+}$ (or Fe$^{2+}$) can be computed by adding (or subtracting) $\Delta E$ to (or from) the free energy surface of state Fe$^{2+}$ (or Fe$^{3+}$). 
%
%
%Additionally, 
On the other hand, under the linear response approximation,\cite{Tateyama_JCP_ET1,ensing_lumiflavin_ET_JCTC}
%
%the following identities
%
we can compute the reorganization free energy as,
\begin{equation}
\label{lambda}
\lambda = (\left \langle \Delta E  \right \rangle _{{{\text {Fe}}^{2+}}} - \left \langle \Delta E  \right \rangle _{{{\text {Fe}}^{3+}}})/2 \enspace . 
\end{equation}
Further, the redox potential for the free energy change during the oxidation reaction can be computed as,
\begin{equation}
\label{deltaf}
\Delta F = (\left \langle \Delta E  \right \rangle _{{{\text {Fe}}^{2+}}} + \left \langle \Delta E  \right \rangle _{{{\text {Fe}}^{3+}}})/2 \enspace ,
\end{equation}
%
%where $\lambda$ is the reorganization free energy and $\Delta F$ is the free energy change during the oxidation reaction (\cref{reaction}).
%
Using \cref{lambda,deltaf}, $\lambda$ and $\Delta F$ can be computed from the ensemble average of $\Delta E$ using AIMD simulations of 
the solvated Fe$^{2+}$ and Fe$^{3+}$ systems.
%
%{\color {blue}
%Also, due to the parabolic nature of the free energy surfaces $F_{{{\text %{Fe}}^{2+}}}(\Delta E)$ and $F_{{{\text {Fe}}^{3+}}}(\Delta E)$, under the %linear response approximation, have identical curvature and the %reorganization free energies computed in both oxidation state should be %equal:
%%
%\begin{equation}
%\label{lambda_eq}
%   \lambda =  \lambda_{{\text {Fe}}^{2+}} = \lambda_{{\text {Fe}}^{3+}} %\enspace .
%\end{equation}
%%
%\cref{lambda_eq} will be used in the \nameref{results_discussion} section  to check the reliability of the linear response assumption for the aqueous ferrous-ferric redox system.
%}

\subsection*{Theory of Hybrid Functional based AIMD Simulations using NSMD}

Hybrid functional based KS-DFT calculation requires the computation of the orbital-dependent Hartree-Fock (HF) exchange energy\cite{Martin-book}
\begin{equation}
\label{e:xc2}
E^{\text {HF}}_{\text {X}} = - \sum_{i,j}^{N_{\text {orb}}}
\left \langle \psi _{i} \left | v_{ij}({\mathbf {r}}_{1}) \right | \psi _{j}
\right \rangle \enspace,
\end{equation}
with
\begin{equation}  
\label{e:vij}
v_{ij}(\mathbf {r}_{1})=\left \langle\psi _{j} \left | \frac{1}{r_{12}} \right | \psi _{i}\right \rangle
\end{equation}
and $r_{12}=\left | \mathbf r_1 - \mathbf r_2 \right | $. 
For computational efficiency, $v_{ij}(\mathbf{r})$ is often computed in reciprocal space (or G-space) using the fast Fourier transform algorithm.\cite{JCP_HFX_Voth,PRB_Car_Wannier}
If $N_{\text {orb}}$ and $N_{\text G}$ are the number of orbitals and the number of
 G-space grid points, respectively, the total computational cost for the HF exchange energy evaluation scales as $N_{\text  {orb}}^2 N_{\text G} \log N_{\text G}$.\cite{JCP_HFX_Voth}
As a consequence, hybrid functional and PWs based AIMD simulations require very high computational time
for typical systems of our interest containing few hundreds of atoms.

In the NSMD method,\cite{JCP_sagar} we reduce the computational cost of HF exchange energy evaluation significantly by screening 
the KS orbital pairs in \cref{e:xc2}.
In particular, we employ an unitary transformation based on the selected columns of the density matrix (SCDM) approach\cite{SCDM_main} to localize the KS orbitals in real space, as
\begin{equation} 
\label{unit}
|\phi _{k} \rangle=\sum_{i}^{N_{\text {orb}}} |\psi _{i} \rangle u_{ik}  \enspace,
\end{equation}
where $u_{ik} \equiv \left ( \mathbf U \right )_{ik}$ and $\mathbf U$ is the unitary matrix.
Now, $E_{\textrm{X}}^{\textrm{HF}}$ can be rewritten using the SCDM based localized orbitals as
\begin{equation} 
\label{e:ehfx:local}
E^{\textrm{HF}}_{\textrm{X}} = - \sum_{i,j}^{N_{\text {orb}}}
\left \langle \phi _{i} \left | v_{ij}({\mathbf {r}}_{1}) \right | \phi _{j}
\right \rangle \enspace,
\end{equation}
where $v_{ij}(\mathbf r)$ is computed as per Eq.~(\ref{e:vij}) using  $\{\phi_i\}$.
For  
non-overlapping orbital pairs in 
real space, $v_{ij}(\textbf{r})$ will be zero  and such pairs of orbitals will not contribute to the HF exchange energy. 
In our computations, we consider 
%employed a pair density cutoff $\rho_{\text {cut}}$ to screen the orbital pairs entering in  Eq.~(\ref{e:ehfx:local}); 
an orbital pair $i$-$j$ only if the criteria
\[ \int d\textbf{r} \left | \phi_{i}(\textbf{r}) \phi_{j}^{*}(\textbf{r}) \right | \geqslant \rho_{\textrm {cut}} \enspace\]
is satisfied.
This screening procedure (based on a pair density cutoff $\rho_{\text {cut}}$) allows for a substantial decrease in the number of orbital pairs entering in  Eq.~(\ref{e:ehfx:local})
during the HF exchange energy computation.
%through neglecting orbital
%products with insignificant contribution,
The computational cost scales now as
$\Tilde{N}_{\text {orb}}^2 N_{\text G} \log N_{\text G}$, where $\Tilde N_{\text {orb}}$ 
is the effective number of orbital products considered in the calculation
after screening, and in our case $\Tilde N_{\text {orb}} \ll N_{\text {orb}}$.

However, during the molecular dynamics simulations, screening of orbital pairs based on $\rho_{\text {cut}}$ introduce small errors in the wavefunctions and nuclear forces resulting in unstable dynamics with a drift in the total energy.
To overcome this, following the work of
K\"uhne {\em et al},\cite{PRL_Kuhne} we employ the NSMD approach
 based on a noise stabilization of the dynamics by coupling a Langevin thermostat to the system. \cite{JCP_sagar}
In this case, the
equations of motion for the ions can be written as
\begin{equation}
\label{lang}
{M}_I {\ddot{\textbf{R}}}_I = \textbf{F}_I - \gamma \dot{\textbf{R}}_I +  {\textbf{x}}_I  \enspace, \enspace I=1,\cdots, N \enspace , 
\end{equation}
where ${M}_I$ is the ionic mass, 
$\textbf{F}_I$ is the
%
%approximated
%
ionic force, 
and $\gamma$ is the friction coefficient of the Langevin thermostat.
${\textbf{x}}_I$ is the random noise that obeys the following expression:
\begin{equation}
\label{lang1}
\left \langle {\textbf x}_I(0){\textbf x}_I(t) \right \rangle=6 \gamma M_I k_{\text B} T \Delta t \enspace,
\end{equation}
where $T$ and $\Delta t$ are the target temperature of the system and the MD time step.
The parameter ${\gamma}$ is chosen in such a way that the correct average temperature is obtained, and the average of the total energy has no drift.

\subsection*{Computational Details}

We modelled a single Fe$^{2+}$ or Fe$^{3+}$ ion solvated in 64 water molecules inside a periodic simulation box of dimension 12.414~{\AA}$\times$12.414~{\AA}$\times$12.414~{\AA}, 
corresponding to a bulk water density of $\sim$1.0 g~cm$^{\text {-3}}$.
No counter ions were included in the system.
Instead, a homogeneous positive background charge was added to maintain charge neutrality.
In our calculations, Fe$^{2+}$ and Fe$^{3+}$ are considered in their high spin configurations.
%with 4 unpaired electron (spin multiplicity 5), 
%
%whereas Fe$^{3+}$ is also in the high spin configuration (3d$^5$) with 5 unpaired electron (spin multiplicity 6).
%
To simulate these open-shell systems, spin polarized periodic DFT calculations were performed.
All computations were performed using 
%periodic
%DFT, employing
% with its pseudopotential implementation~\cite{marx-hutter-book}
a modified version of the CPMD program,\cite{cpmd,KLOFFEL2021} where we implemented the
%noise stabilized molecular dynamics (
NSMD approach.\cite{JCP_sagar}
We employed both BLYP\cite{PRA_GGA_Becke,PRB_GGA_LYP} (GGA) and B3LYP\cite{JCP_B3LYP} (hybrid) exchange correlation functionals together with norm-conserving Troullier-Martin type pseudopotentials.\cite{PRB_TM}  
The pseudopotential for iron includes the non-linear core correction.\cite{nlcc}
The wavefunctions were expanded in a PW basis set with a cutoff of 90~Ry.
Born-Oppenheimer molecular dynamics (BOMD) simulations were carried out in the canonical (NVT) ensemble 
%using the Nos{\'e}--Hoover chain thermostat\cite{NHC}
at 300~K with a time step of $0.48$~fs.
%

%Two sets of simulations were performed for solvated Fe$^{2+}$ and Fe$^{3+}$ systems: 
%
%
%(a) {\bfseries BLYP}: 
In the BLYP (GGA) functional based $NVT$ ensemble simulations, 
%We set the 
%, BOMD simulation using the BLYP (GGA) functional with 
%tightly converged wavefunctions (i.e., till the maximum 
%In SCF calculations, 
the wavefunctions were optimized every MD step till the wavefunction gradient fell below 
%criteria in SCF such that wavefunction gradient is converged below
%wavefunction gradient turned less than 
10$^{-6}$ a.u.
%in the BOMD simulations.
%
A Nos{\'e}--Hoover chain thermostat\cite{NHC} was used to control the temperature of the system at 300~K.
The always stable predictor corrector (ASPC) extrapolation scheme\cite{JCC_ASPC} of order 5 was used to obtain the initial guess of wavefunctions.
On the other hand, for all $NVT$ MD simulations using the B3LYP (hybrid) functional,
%In SCF calculations, 
the wavefunctions were optimized till 
the magnitude of change in energy is below 1$\times$10$^{-3}$~a.u.
%the energy change in successive self consistent field iterations dropped below 1$\times$10$^{-3}$~a.u.}
%wavefunction was optimized till {\color {red} the wavefunction gradient dropped below 1$\times$10$^{-3}$~a.u.}
%
%Here we used, BOMD simulation using B3LYP (hybrid) functional employing the NSMD approach.
%
%NSMD allows to speed up computationally demanding hybrid functional based BOMD simulations (more details about the NSMD method can be found in Ref.~\citenum{JCP_sagar}).
%
%In this scheme, we used the SCDM\cite{SCDM_main} orbitals to screen the non-overlapping orbital pairs during the computation of the HF exchange energy.
%and all the non overlapping orbital pairs were excluded from the exchange energy computation.
%
%In our calculations, 
Exchange integrals were calculated by the screened SCDM orbitals with a pair density cutoff of $\rho_{\text {cut}} = 4\times 10^{-3}$.  
%
%
%With this screening approach, the noise in the wavefunction gradients was such that 
%achieving a tight convergence 
%was practically not possible. 
%
%till its maximum gradient attaining {\color {red} (change in total energy in the successive iteration is less than 10$^{-3}$~a.u.)} magnitude less than 10$^{-3}$~a.u.
%
%Thus, we converged the 
%
We employed the NSMD approach to perform BOMD simulations.
The Langevin thermostat parameter, the friction coefficient ${\gamma}$, was set to 1$\times$10$^{-3}$ a.u.$^{-1}$, and the target temperature $T$ was set to 300~K.
For the initial guess of the wavefunctions, the ASPC extrapolation scheme of order 2 was used.

For these systems, 
the hybrid calculations using NSMD could achieve a speed up of $\sim$12~times per MD step compared to the conventional hybrid density functional based AIMD simulation
on 280 cores (14~nodes, each with two Intel Xeon E5-2630v4 Broadwell chips with 10 cores running at 2.2 GHz with 64 GB of RAM).

%Thus, we performed four independent simualations: 
%(i) Fe$^{2+}$ with GGA (BLYP) functional,
%(ii) Fe$^{2+}$ with hybrid (B3LYP) functional,
%(iii) Fe$^{3+}$ with GGA (BLYP) functional, and
%(iv) Fe$^{3+}$ with hybrid (B3LYP) functional.
%Fe$^{2+}$/Fe$^{3+}$ systems with both hybrid (B3LYP) and GGA (BLYP) functionals.
%
We preformed  20~ps (10~ps) of equilibration for the {B3LYP} (BLYP) simulations. %respectively.
Subsequent to this, $\sim$20 ps long production simulations were carried out in all the cases.
%were generated for all simulations which were used in our analysis.
%
%The total length of the simulations were XX ps.
%
%We discarded initial 10 ps of the trajectories, and considered the last 30 ps of the trajectories for our analysis.
%
To calculate the vertical energy gap (\cref{gap}), we sampled equidistant configurations from the trajectories with a time interval of 20~fs.  
%
% was calculated for snapshots with time At an interval of 20 fs.
%
In total, we took 1000 equally sampled configurations from each of the trajectories.
For every configuration, single point energy calculations were performed for the reduced (Fe$^{2+}$) and the oxidized (Fe$^{3+}$) states.
%different number of electrons and charge .
%
%During the single point calculations, the wavefunction convergence criteria was set to 10$^{-6}$ a.u. for the {\bfseries BLYP} runs.
%
%While, due to the poor convergence during the SCF iterations, we used a wavefunction convergence criteria of 10$^{-5}$ a.u. for the {\bfseries B3LYP} runs.

Additionally, to estimate the finite size effects, we performed {BLYP} calculations with two larger unit cells: (a) one Fe$^{2+}$/Fe$^{3+}$ ion solvated in 92 water molecules within a periodic cubic simulation box of side 14.010~{\AA}; (b) one Fe$^{2+}$/Fe$^{3+}$ ion solvated in 137 water molecules inside a periodic cubic simulation box of side 16.000~{\AA}.
%with increased number of water molecules at the level of BLYP (GGA) functional.
%
%For all these systems, the simulation parameters were identical to that of {\bfseries BLYP} simulations.
%
%The first system contains one Fe$^{2+}$/Fe$^{3+}$ ion solvated in 92 water molecules inside a periodic simulation box of dimension 14.01~{\AA}$\times$14.01~{\AA}$\times$14.01~{\AA}.
%
For these two systems, we performed 5~ps $NVT$  equilibration and 7~ps $NVT$ production runs.
%trajectories for our analysis.
%
The vertical energy gap was calculated by taking 500~snapshots sampled every $\sim$14.5 fs.
%
%In total, we have {\color {blue} X} snapshots for each of the trajectories.
%
%
%
%The second system contains one Fe$^{2+}$/Fe$^{3+}$ ion solvated in 137 water molecules inside a periodic simulation box of dimension 16.0~{\AA}$\times$16.0~{\AA}$\times$16.0~{\AA}.
%
%In these cases, we carried out a total of 12 ps long simulations and collected the data for our analysis from the last 7 ps trajectories.
%
%The vertical energy gap was calculated for 500 snapshots with time interval of $\sim$14.5 fs.
%

%\clearpage

%%%%%%%%%%%%%%%%%%%%%%%%%%%%%%%

\section*{Results and Discussion}
\label{results_discussion}

\subsection*{Solvation Structure}

%\subsubsection{Radial pair distribution function}

%
\begin{figure*}[h]
\begin{center}
\includegraphics[scale=0.25]{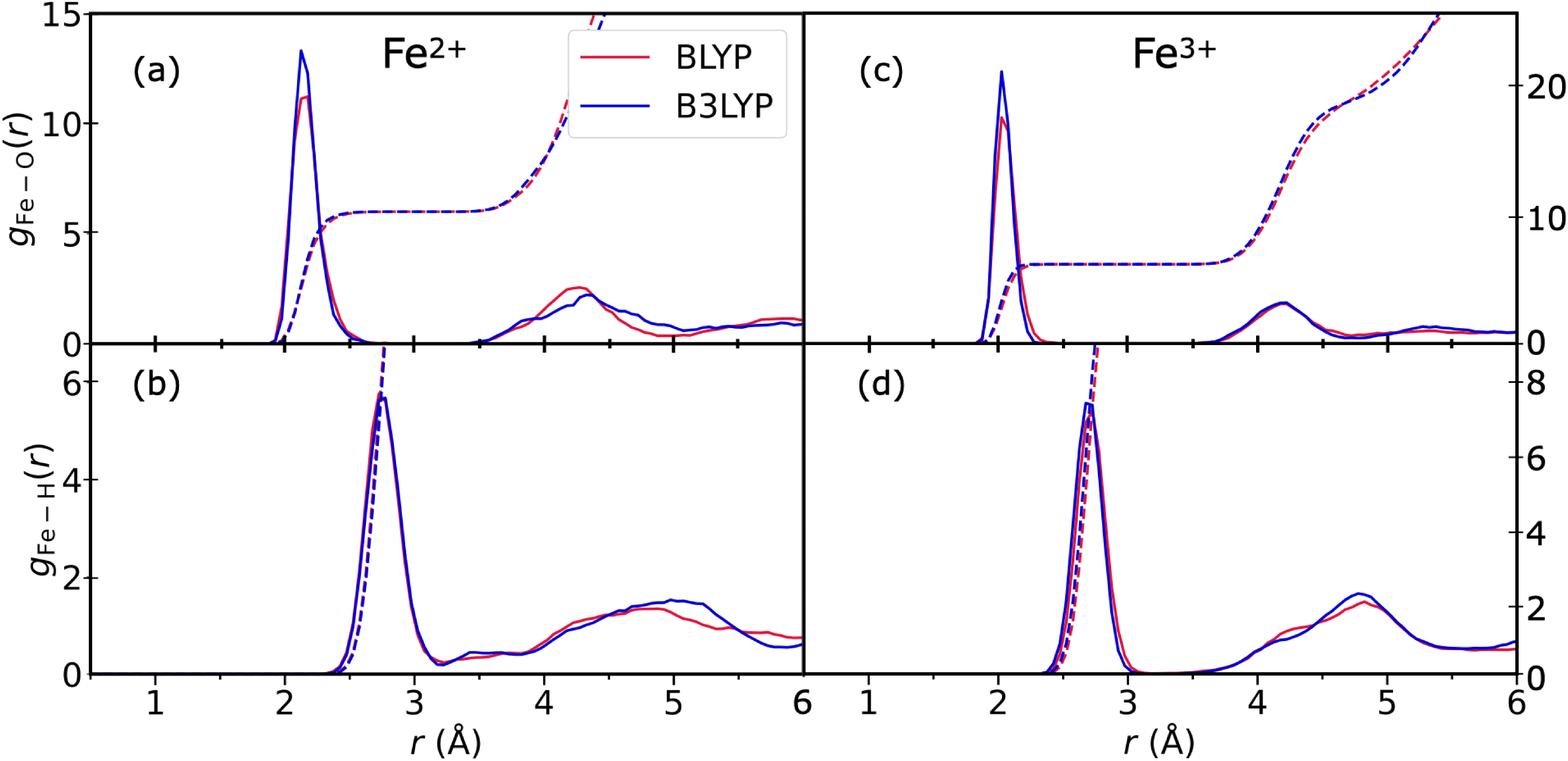}% Here is how to import EPS art
\caption{Comparison of (a) Fe$^{2+}$--O, (b) Fe$^{2+}$--H, (c) Fe$^{3+}$--O and (d) Fe$^{3+}$--H radial distribution functions (RDFs) for {BLYP} and {B3LYP}.
The running integration number of the RDFs are also shown with dashed lines.
}
 \label{g_of_r_fe} 
 \end{center}
\end{figure*}

\begin{table*}
\begin{center}
\caption{Comparison of structural properties of the first and second solvation shell (SS) of water around Fe$^{2+}$ and Fe$^{3+}$ ions in our simulations with other data from simulations (Sim.) and experimental (Expt.) studies.}\label{tab1}
\begin{tabular}{ll|cccc|cccc}
\toprule
\multirow{2}{*}{SS} & \multirow{2}{*}{Property} & \multicolumn{4}{c|}{Fe$^{2+}$} & \multicolumn{4}{c}{Fe$^{3+}$} \\
               &                    & BLYP & B3LYP & Sim.\textsuperscript{[a]} & Expt.\textsuperscript{[b]}   & BLYP & B3LYP  & Sim.\textsuperscript{[a]}  & Expt.\textsuperscript{[b]}   \\
\midrule
%1                  & 2                  & 3   & 4   & 5   & 6   & 7   & 8   & 9   & 0   \\
& {$r_{\text {Fe--O}}^{\text {max}}$ (\AA)} &   2.17   &  2.15 & 2.09 -- 2.13 & 2.10 -- 2.28  &  2.04  &   2.03  & 1.96 -- 2.10 & 1.98 -- 2.05  \\
  & {$r_{\text {Fe--H}}^{\text {max}}$ (\AA)} &   2.73  &    2.75    & 2.40 -- 2.76 &  &    2.71    &   2.70    & 2.66 -- 2.77 &   \\ 
1st & {CN} &     6.0    &    6.0    & 6.0 & 6.0  &    6.0    &   6.0    & 6.0 &  6.0   \\ 
 & {$r_{\text {O--H}}$ (\AA)} &   0.99    &    0.98    &  1.03 &  &   1.0   &   1.0    &  1.01 -- 1.06 & \\
 & {$\theta_{\text {H--O--H}}$ (\textdegree)} & 106.2   &   106.3 & 102.0   &  & 107.1  &  107.1   & 98.8 -- 107.0 &  \\
\midrule
  & {$r_{\text {Fe--O}}^{\text {max}}$ (\AA)} &  4.29   &   4.35    & 4.25 -- 4.50 & 4.30 -- 4.51 &  4.20   &   4.20   & 4.11 -- 4.30 & 4.09 -- 4.80 \\
  & {$r_{\text {Fe--H}}^{\text {max}}$ (\AA)} &  4.81   &   4.99    &  &  &   4.81   &  4.77     & 4.76 -- 4.96 &\\
2nd  & {CN} &     13.0    &    15.3    &  11.6 -- 14.4  & 12  &  12.2    &   12.6     & 11.0 -- 14.0 & 12  \\
 & {$r_{\text {O--H}}$ (\AA)} &     0.99   &    0.98    &  1.02 & &  0.99  & 0.98    & 1.02 &  \\
 & {$\theta_{\text {H--O--H}}$ (\textdegree)} & 106.4    &  107.0    & 104.7 & &  106.7  &  105.9    & 104.4 & \\
\bottomrule	
\end{tabular}
\end{center}
\footnotesize{\textsf{[a] Refs.~\citenum{theory_cpmd_fe3_1,theory_fe2_fe3_2,theory_qmmm_fe2_fe3_3,theory_fe2_4,theory_fe2_fe3_5,theory_fe3_6,theory_fe2_fe3_7,theory_fe2_8,theory_fe2_9,theory_fe2_fe3_10}. 
[b] Refs.~\citenum{expt_fe2_fe3_1,expt_fe2_fe3_2,expt_fe2_fe3_3,expt_fe3_4}. 
}}
\end{table*}

First, we compare the solvation structure of the Fe$^{2+/3+}$ ions with {BLYP} and {B3LYP} functionals.
The radial pair distribution functions (RDFs) of the metal with oxygen and hydrogen were calculated from all the trajectories and are shown in \cref{g_of_r_fe}.
In all cases, we observe two distinct solvation shells around the Fe$^{2+/3+}$ ions.
The first (second) peak of the Fe$^{2+}$--O distribution appears at 2.17 (4.29) {\AA} and 2.15 (4.35) {\AA} with BLYP and B3LYP functionals, respectively.
%
%The next peak due to the second solvation shell of the Fe$^{2+}$ ion appears at 4.29 {\AA} and 4.35 {\AA} for BLYP and B3LYP.
%
%
%
In case of the Fe$^{3+}$--O distributions, the first and the second peaks appear at nearly the same location for both functionals. 
%\gga and \hyb (respectively), while the second peak is at the same point (4.20~{\AA}).
%for both BLYP and B3LYP.
%
%
The first and the second peaks of the Fe$^{2+}$--H distributions are 0.02~{\AA} and 0.18~{\AA} shorter for BLYP compared to B3LYP.
On the other hand,  the locations of the first and second peaks 
in the Fe$^{3+}$--H distributions
are not having much effect on the choice of the functional.
%the same with the two functionals.
%BLYP compared to B3LYP.
%
These results are consistent with the previously reported computational\cite{theory_cpmd_fe3_1,theory_fe2_fe3_2,theory_qmmm_fe2_fe3_3,theory_fe2_4,theory_fe2_fe3_5,theory_fe3_6,theory_fe2_fe3_7,theory_fe2_8,theory_fe2_9,theory_fe2_fe3_10}
as well as experimental\cite{expt_fe2_fe3_1,expt_fe2_fe3_2,expt_fe2_fe3_3,expt_fe3_4} data; see also \cref{tab1}.
%{\tt \color{red} <<Why peak is different for Fe2+??>> }
%

In all cases, the RDFs are
%become zero in the range $\sim$ 2.5--3.5 {\AA},
indicating a clear separation of the first and second solvation shell.
During our simulations, we did not observe any exchange of water molecules between the first and second solvation shell.
Due to the higher charge of the Fe$^{3+}$ ion compared to its reduced form, we can notice that the first and the second peaks of the Fe$^{3+}$--O distributions are more sharply peaked as compared to the Fe$^{2+}$ case.
%
%

%{\color{red} The difference in the position of the first peak for the Fe$^{2+}$--O and Fe$^{3+}$--O distributions is 0.13 {\AA} with the BLYP functional, whereas the same difference is 0.12 {\AA} with B3LYP.
%
%For the same distributions, the difference in the position of the second peak is 0.09 {\AA} with BLYP as compared to a value of 0.15 {\AA} with B3LYP.
%
%Although the location of the first peak of the Fe$^{2+/3+}$--O distribution is almost similar for both functionals, the distribution is sharper in the B3LYP case.
%
%
%These observations suggest that the effect of the functional is minimal for the first solvation shell, whereas, we observe small difference in the second solvation shell structure.}
%{\tt \color{red} <<THESE ARE REPEATED INFORMATION. CHECK AND DELETE LINES IN RED!!!!!>>}

\begin{figure}[h]
\begin{center}
\includegraphics[scale=0.2]{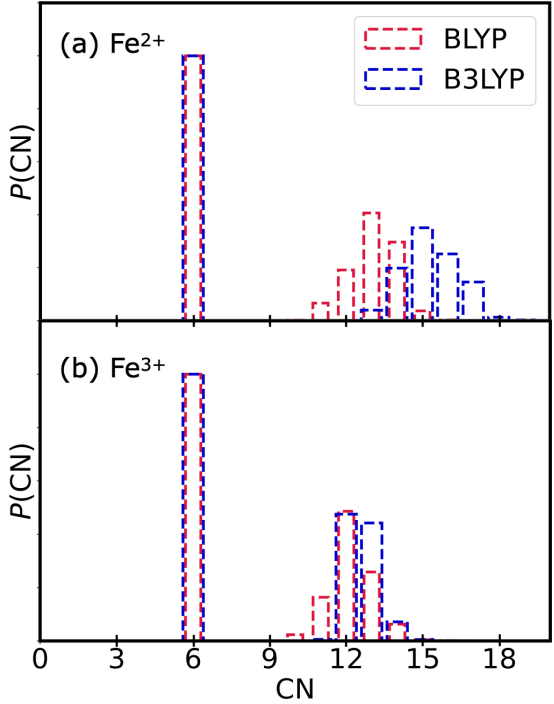}% Here is how to import EPS art
\caption{Coordination number (CN) distribution in the first and second solvation shells of (a) Fe$^{2+}$ and (b) Fe$^{3+}$ ions.}
\label{fig_cn} 
\end{center}
\end{figure}

\begin{figure}[h]
\begin{center}
\includegraphics[scale=0.2]{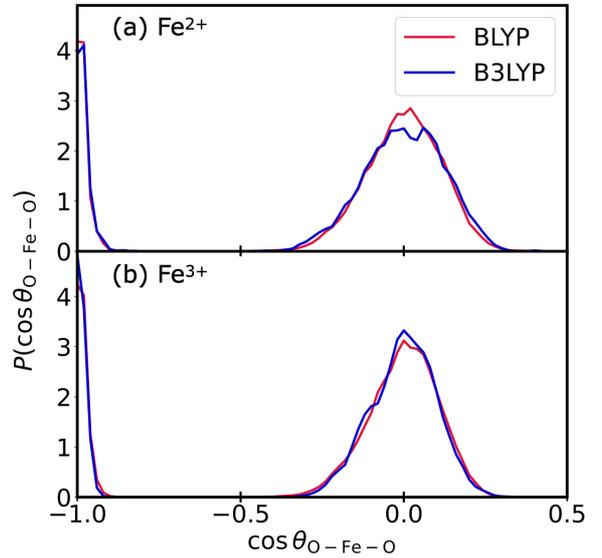}% Here is how to import EPS art
\caption{Comparison of the probability distribution of the cosine of O--Fe--O angle in the first solvation shell of (a) Fe$^{2+}$ and (b) Fe$^{3+}$ ions.}
\label{fig_adf} 
\end{center}
\end{figure}

%\subsubsection{Coordination number distribution}
Coordination numbers (CNs) of the ions with water molecules are 
computed by integrating RDFs; see \cref{fig_cn}.
%we also showed the 
%Running integration number of RDFs, which indicates
%
%the coordination numbers (CNs) of the ions up to any radial distance.
%can be obtained by integrating the distribution functions upto the first and second minimum for calculating the fist and second shell CNs, respectively.
%
%This number tells the number of water molecules present in the first or second solvation shell.
%
%
The most preferred CN for both ions in water is 6.0 (see also \cref{tab1}), which is in agreement with earlier studies.
%
%Thus, we conclude that the Fe$^{2+/3+}$ ions 
%are surrounded by 6 water molecules in the first solvation shell and 
%form a stable hexaaqua complex in aqueous medium in agreement with earlier studies.
%
%The CN of the second solvation shell can be found by subtracting the first shell CN from the value of the running integration number at the second minima of the Fe$^{2+/3+}$--O distributions.
%
%In our analysis, the second minima were considered at 4.9 {\AA} and 5.1 {\AA} for the Fe$^{2+}$ case with BLYP and B3LYP, respectively,
%
%while the second minima were considered at 4.7 {\AA} and 4.8 {\AA} for the Fe$^{3+}$ case with BLYP and B3LYP.
%
%
%The computed CNs for the first and second solvation shells are tabulated in \cref{tab1}.
%
We computed the CNs for the second solvation shell, and we find that our predicted second shell
CN is larger than the experimental data, but close to some of the computational studies.
The computed CN of the second shell for Fe$^{2+}$ is higher as compared to that of Fe$^{3+}$.
%
%This is because the higher charge of Fe$^{3+}$ maintains a well defined solvation structure which is beyond the first shell.
%%
%This make the second peak of RDFs sharper and contains less waters.
%
%Whereas, for the Fe$^{2+}$ case, the solvent molecules are more mobile and the peak is broader.
%
%Thus contains more molecules in the second solvation shell.
%
Interestingly, in the {B3LYP} simulations, we observe higher CN values for the second shell
%computed by integrating RDFs; 
compared to BLYP,
see \cref{fig_cn}.

%compared to {\bfseries BLYP}, and is more profound for the Fe$^{2+}$ case; see \cref{fig_cn}.
%

To further examine the geometry of the hexaaqua complex, we compare in \cref{fig_adf}  the probability distribution of the cosine of O--Fe--O angle in the first solvation shell of the Fe$^{2+/3+}$ ions from the BLYP and B3LYP functional based AIMD simulations.
 The distribution functions show two peaks at $-1.0$ and $0.0$, 
indicating almost perfect octahedral arrangement of the water molecules. %{\tt \color{red} <<change figure>>} 
%
%Also, we have reported in \cref{tab1} the intramolecular geometry of the water molecules in the first and second solvation shell.
%
By comparing the average O--H bond distance and H--O--H angle of the water molecules in the first and the second coordination sphere, 
we conclude that the solvation structure  is less affected by the functionals, see \cref{tab1}.
%for the first solvation shell.
%whereas, slight change is observed in case of second solvation shell. 
%
%For the second solvation shell, we observe minor changes in the intramolecular geometries due to the functionals.

%\clearpage

\subsection*{\label{sec:ET}Electron Transfer Reaction}

\begin{figure}[h]
\begin{center}
\includegraphics[scale=0.18]{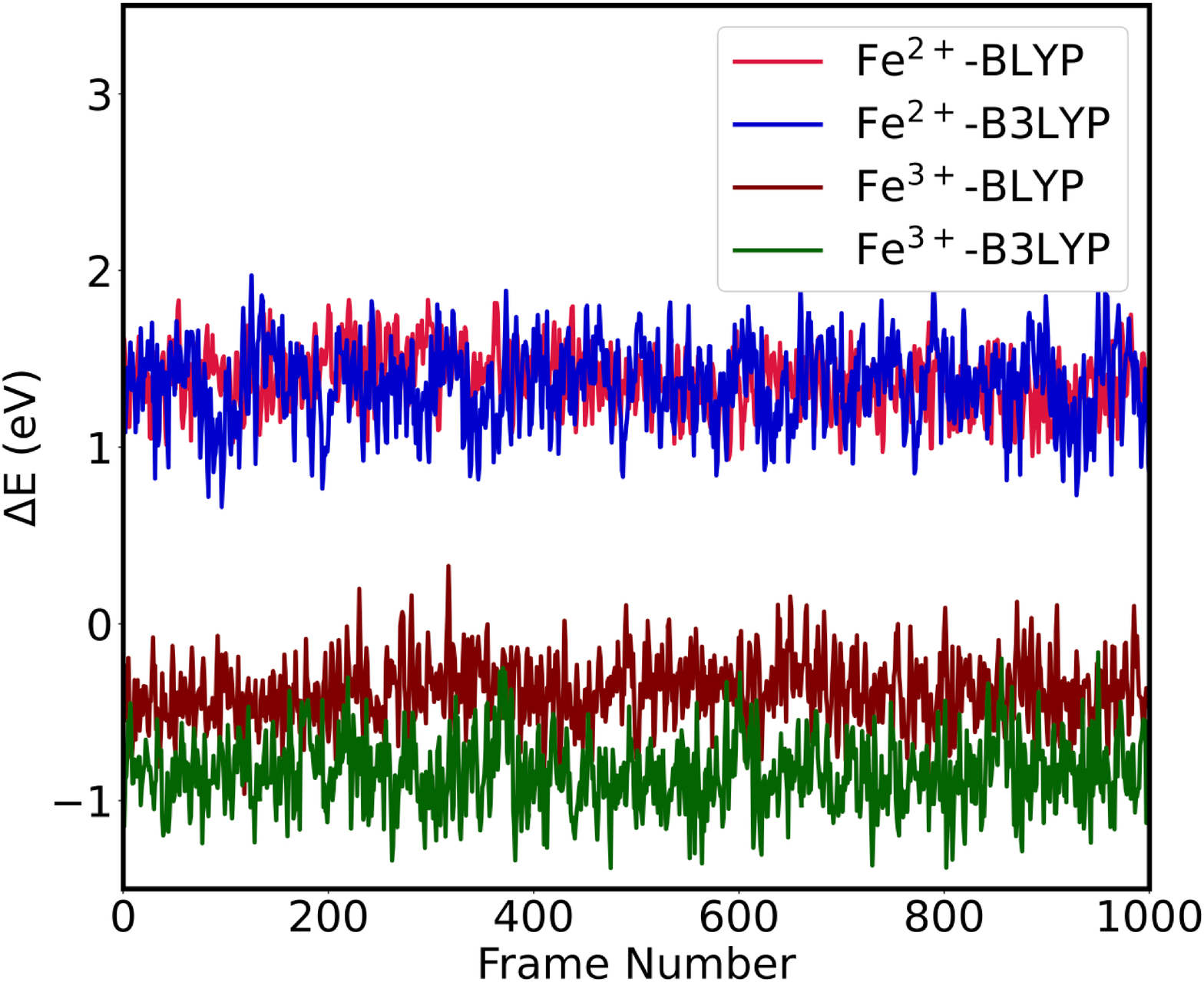}% Here is how to import EPS art
\caption{Vertical energy gap ($\Delta E$) during {BLYP} and {B3LYP} simulations of ${\text {Fe}}^{2+}$ and ${\text{Fe}}^{3+}$ systems.
%(BLYP), ${\text {Fe}}^{2+}$ (B3LYP), ${\text {Fe}}^{3+}$ (BLYP) and ${\text {Fe}}^{3+}$ (B3LYP) systems.
}
\label{deltaE_fluct}
\end{center}
\end{figure}

We then computed the redox properties of the aqueous Fe$^{2+}$ and Fe$^{3+}$ systems.
The vertical energy gap  (\cref{gap}) is calculated for the structures sampled during the MD simulations.
%of the aqueous ${\text {Fe}}^{2+/3+}$ (BLYP) and ${\text {Fe}}^{2+/3+}$ (B3LYP) systems.
%
%The fluctuations of the energy gap are shown in \cref{deltaE_fluct} for all these trajectories.
%
%$\Delta E$ fluctuation is nearly in the s similar range for the Fe$^{2+}$ (BLYP) and Fe$^{2+}$ (B3LYP) systems.
%
%Interestingly, the fluctuation of $\Delta E$ for Fe$^{3+}$ However, the ranges of the fluctuations in $\Delta E$ are clearly separable for both functionals in the case of Fe$^{3+}$ simulations.
%
%section, all
Redox properties were calculated from the fluctuation of $\Delta E$ and are reported in \cref{tab2}.
%
%The statistical errors in our computed properties are also reported based on a block averaging analysis.\cite{leach,ensing_lumiflavin_ET_JCTC}
%
$\left \langle \Delta E  \right \rangle _{{{\text {Fe}}^{2+}}}$, 
computed from {BLYP} and {B3LYP} simulations, are almost identical.
However, $\left \langle \Delta E  \right \rangle _{{{\text {Fe}}^{3+}}}$ from {the BLYP} run is 0.46~eV larger than for 
{B3LYP}. 
%B3LYP.
%
%The standard deviations of the $\Delta E$ fluctuations are 0.17 eV and 0.19 eV for the Fe$^{2+}$ (BLYP) and  Fe$^{3+}$ (BLYP) cases, respectively.
%
Nearly identical values of $\sigma_{\Delta E}^{{\text {Fe}}^{2+}}$ and $\sigma_{\Delta E}^{{\text {Fe}}^{3+}}$
show the validity of the linear response assumption underlying the Marcus theory.
%
%For the Fe$^{2+}$ (B3LYP) and  Fe$^{3+}$ (B3LYP) cases, the standard deviations, $\sigma_{\Delta E}^{{\text {Fe}}^{2+}}$=0.23 eV and $\sigma_{\Delta E}^{{\text {Fe}}^{3+}}$=0.20 eV, are also found to be similar.
%
The standard deviations of $\Delta E$ are higher in {B3LYP} cases as compared to {BLYP}, implying a much larger spread in the $\Delta E$ values during the hybrid functional based simulations.
%
%{\tt \color{red} <<WHY??? NOT REALLY FOR FE3+. ONLY FOR FE2+>>}

\begin{table}
\begin{center}
\caption{Redox properties computed from the AIMD simulation of the aqueous ${\text {Fe}}^{2+}$ and ${\text {Fe}}^{3+}$ systems.
%The statistical errors are measured by a block averaging analysis.
}\label{tab2}
\begin{tabular}{lcc}	
\toprule		
Properties  &   BLYP (eV) & B3LYP (eV) \\
\midrule
$\left \langle \Delta E  \right \rangle _{{{\text {Fe}}^{2+}}}$ & 1.38 $\pm$ 0.03 &  1.35 $\pm$ 0.02 \\
$\left \langle \Delta E  \right \rangle _{{{\text {Fe}}^{3+}}}$ & -0.38 $\pm$ 0.02  & -0.84 $\pm$ 0.02 \\ % pm not updated
$\sigma_{\Delta E}^{{\text {Fe}}^{2+}}$ & 0.17 &  0.23 \\
$\sigma_{\Delta E}^{{\text {Fe}}^{3+}}$ & 0.19  & 0.20 \\
$\lambda_{{\text {Fe}}^{2+}}$ &  0.87  &  1.07  \\  % not updated
$\lambda_{{\text {Fe}}^{3+}}$ &  0.87  &  1.07  \\
$\lambda$  & 0.88 $\pm$ 0.02 &  1.10 $\pm$ 0.01 \\ 
$\Delta F$  & 0.50 $\pm$ 0.02  & 0.26 $\pm$ 0.01 \\
\bottomrule	
\end{tabular}
\end{center}
\end{table}

\begin{figure}[h]
\begin{center}
\includegraphics[scale=0.18]{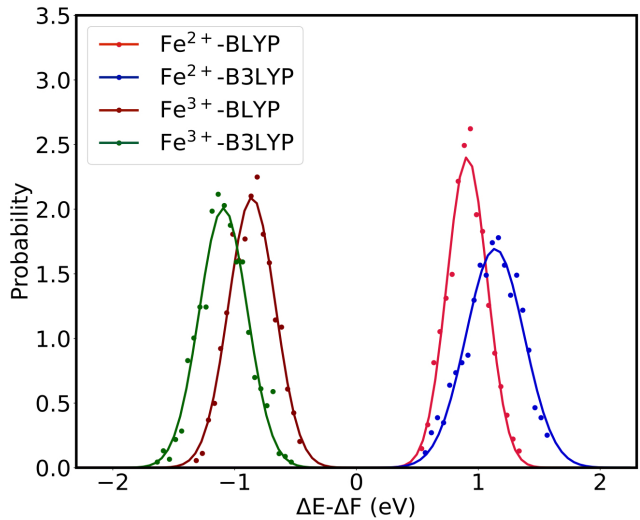}% Here is how to import EPS art
\caption{Probability distribution of $\Delta E$ for the aqueous ${\text {Fe}}^{2+}$ and ${\text {Fe}}^{3+}$ systems. The gap energies are shifted by $\Delta F$ (\cref{deltaf}). 
Probability distributions are fitted with Gaussian functions (solid lines).}
\label{prob_deltaE}
\end{center}
\end{figure}

\begin{figure}[h]
\begin{center}
\includegraphics[scale=0.18]{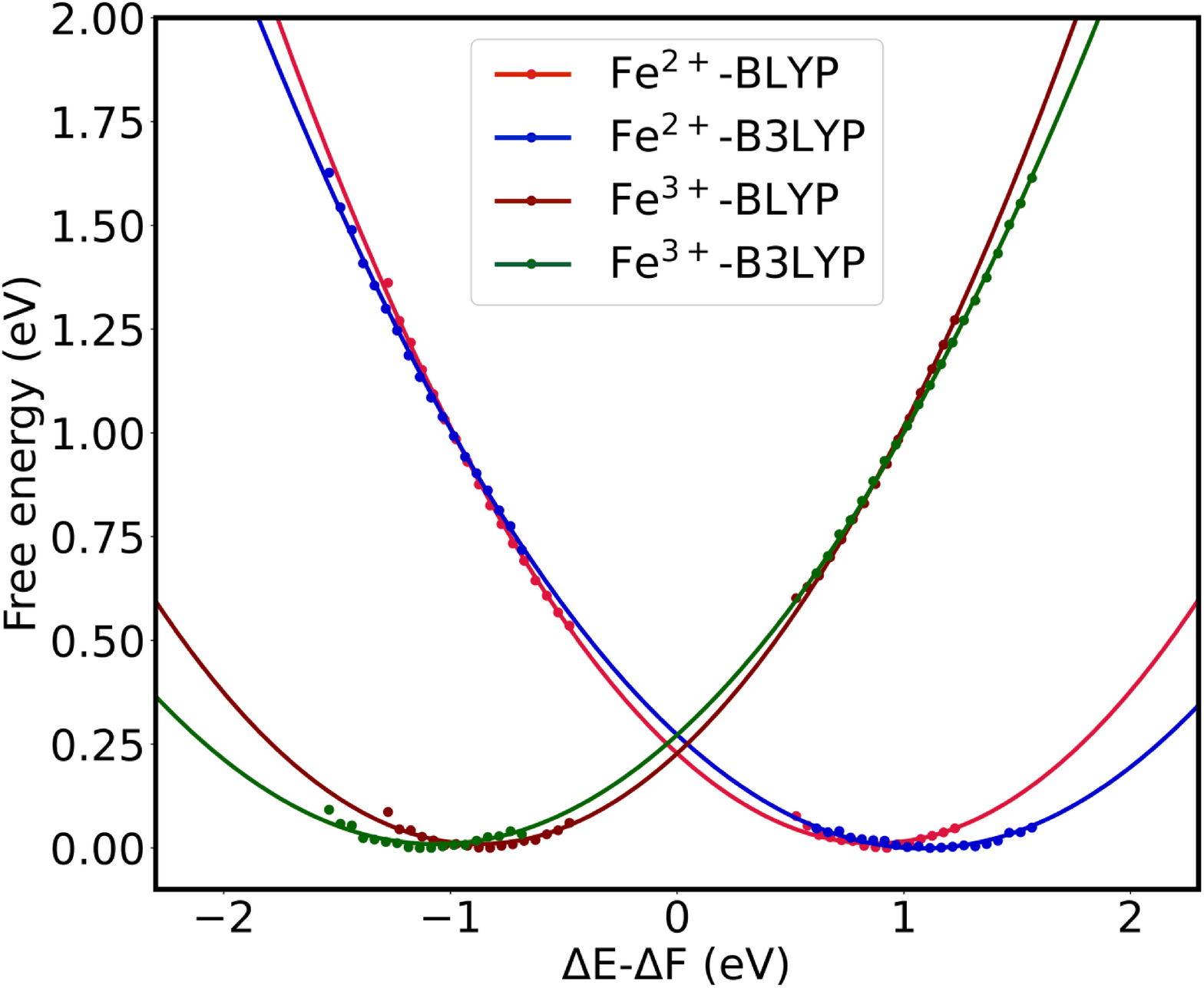}% Here is how to import EPS art
\caption{Free energy profiles as a function of $\Delta E$ for the aqueous ${\text {Fe}}^{2+}$ and ${\text {Fe}}^{3+}$ systems.
%Gap energies are shifted by $\Delta F$ (\cref{deltaf}).
Free energy curves are fitted with parabolic functions (solid lines).
%The errors in the fits are indicated with $R^2$ values.
%The free energy relation of \cref{lineqn} is used to generate the data points at the non equilibrium positions.
}
\label{fes_deltaE}
\end{center}
\end{figure}

In \cref{prob_deltaE}, we show the probability distributions of $\Delta E$.
%
%
%For these plots, we have only considered data points that fall inside the range from 
%$\left \langle \Delta E \right \rangle_{M} - 2\sigma _{\Delta E}^{M}$ to $\left \langle \Delta E \right \rangle_{M} + 2\sigma _{\Delta E}^{M}$.
%
%A bin size of 0.05 eV was used to collect the data points.
%
Gaussian functions are used to fit these probability distributions within the range [$\left \langle \Delta E \right \rangle_{M} - 2\sigma _{\Delta E}^{M}$, $\left \langle \Delta E \right \rangle_{M} + 2\sigma _{\Delta E}^{M}$].
%
%
%The distributions are fitted quite well with the Gaussian functions with
%
The $R^2$ values of these fits are 0.973, 0.952, 0.996 and 0.962 for the ${\text {Fe}}^{2+}$ (BLYP), ${\text {Fe}}^{2+}$ (B3LYP), ${\text {Fe}}^{3+}$ (BLYP) and ${\text {Fe}}^{3+}$ (B3LYP) systems, respectively.
We notice a broader distribution in case of {B3LYP} results as compared to {BLYP}, in accordance to the values of 
the standard deviations we computed earlier.
The consequence of this observation in the curvature of the free energy profiles will be discussed shortly.
Free energy surfaces are reconstructed from the probability distributions (using \cref{fes_con}), see \cref{fes_deltaE}.
We have aligned the minima of all free energy curves to zero.
%
%Similar to the $ P_{{M}} (\Delta E)$ plot in \cref{prob_deltaE}, only values within the range of $\left \langle \Delta E \right \rangle_{M} - 2\sigma _{\Delta E}^{M}$ to $\left \langle \Delta E \right \rangle_{M} + 2\sigma _{\Delta E}^{M}$ were considered for the equilibrium region of the free energy curves.
%
For the region far from equilibrium, we used the linear free energy relation of \cref{lineqn} to obtain the data points.
Free energy data were then fitted with parabolic functions, and the 
%
%
%We observe that the free energy profiles are well approximated with parabolic functions with 
$R^2$ values for the fits are 0.9995, 0.9998, 0.9995 and 0.9998 for the ${\text {Fe}}^{2+}$ (BLYP), ${\text {Fe}}^{2+}$ (B3LYP), ${\text {Fe}}^{3+}$ (BLYP) and ${\text {Fe}}^{3+}$ (B3LYP) cases, respectively. 
This result suggests that the aqueous ${\text {Fe}}^{2+/3+}$ redox system behaves linearly with respect to the solvent response, as reported in earlier studies.\cite{ET_Fe_1,ET_Fe_4_Sit,ET_Fe_5,ET_Fe_6}
Interestingly, the curvature and the location of the minima in the free energy profiles differ for the two functionals.
%
%{\tt \color{red} <<WHY???>> }

Finally, the solvent reorganization free energies were calculated from the fitted parabolic curves (\cref{lambda2,lambda3}) and 
from the vertical energy data (\cref{lambda}).
%
%The results are tabulated for both functionals in \cref{tab2}.
%
%Additionally, the reorganization energy ($\lambda$) was calculated from \cref{lambda}.
%
We find that the computed reorganization energies agree well with each other.
This further validates the linear response assumption of Marcus theory.
%
%The value of $\lambda$ is 0.88~eV and 1.10~eV using BLYP and B3LYP functionals, respectively.
%
We find that the B3LYP functional based calculations predict a higher $\lambda$ value (1.10~eV) compared to GGA (0.88~eV). 
%{\tt \color{red} <<Why hybrid gives the higher lambda value?>> }
%
%
%
%We found that although both of the functionals are underestimating the values, the hybrid functional is more closer to the experimental results.
%
Also, we find that the calculated redox potential is 0.50~eV and 0.26~eV for BLYP and B3LYP cases, respectively.

Before comparing our results with the experimental estimates, we need to account for the errors due to the finite size effect. 
%of our simulation cells.
%
It has been reported that the finite size effect\cite{blumberger_Ru_ET_JACS,sprik_M_ET_JPCB,size_effect_sprik,sprik_Ru_ET_JCTC,blumberger_size_effect} contributes substantially to the %of the system introduces a good amount of error in the predicted
reorganization free energy.
Generally, the error in $\lambda$ is inversely proportional to the side length of the box and is found to be as large as an eV or more.\cite{blumberger_Ru_ET_JACS,sprik_M_ET_JPCB}
{
Sprik and co-workers\cite{sprik_M_ET_JPCB} have concluded that $\lambda$ is usually underestimated.
%in a simulation with finite cell size and the missing contribution scales linearly with $1/L$.
%
These authors have shown that correction for the solvent
reorganization energy is 
\begin{equation}
\label{correction}
    \lambda_{\text C} = - \frac{{\Delta q}^2}{2L} \xi_{\text {EW}} \left ( \frac{1}{\epsilon_{\text {op}}} - \frac{1}{\epsilon_{\text {st}}} \right ) \enspace ,
\end{equation}
where $L$, $\Delta q$ and $\xi_{\text {EW}}$ are the length of the periodic simulation cell, change in charge during the oxidation reaction and the Madelung constant ($\xi_{\text {EW}}=-2.837297$ for a cubic unit cell), respectively.
Here, $\epsilon_{\text {op}}$ = 1.78, $\epsilon_{\text {st}}$ = 78.4 are the optical and static dielectric constant, respectively.\cite{lide2008crc,haynes2015crc}
Using this, $\lambda_{\text C}$ for the 64 water system was found to be about 0.91 eV.

\begin{table}
\begin{center}
\caption{Redox properties computed from the AIMD simulation of larger aqueous ${\text {Fe}}^{2+/3+}$ systems using
BLYP density functional.
}\label{tab3}
\begin{tabular}{lcc}	
\toprule		
Properties  &   System 1  & System 2  \\ 
\midrule
$N_{\text {water}}$\textsuperscript{[a]} & 92 & 137 \\
$L$ ({\AA}) \textsuperscript{[b]} & 14.01 &  16.00 \\
$\left \langle \Delta E  \right \rangle _{{{\text {Fe}}^{2+}}}$ (eV) & 1.50 $\pm$ 0.03 &  1.52 $\pm$ 0.02 \\
$\left \langle \Delta E  \right \rangle _{{{\text {Fe}}^{3+}}}$ (eV) & -0.43 $\pm$ 0.02  & -0.47 $\pm$ 0.02 \\ % pm not updated
$\sigma_{\Delta E}^{{\text {Fe}}^{2+}}$ (eV) & 0.18 &  0.16 \\
$\sigma_{\Delta E}^{{\text {Fe}}^{3+}}$ (eV) & 0.20  & 0.20 \\
$\lambda$ (eV) & 0.96 $\pm$ 0.02 &  0.99 $\pm$ 0.01 \\ 
$\Delta F$ (eV) & 0.53 $\pm$ 0.02  & 0.52 $\pm$ 0.01 \\
\bottomrule	
\end{tabular}
\end{center}
\footnotesize{\textsf{[a] Number of water molecules present in the system.
[b] Box length of the simulation cell.
}}
\end{table}

\begin{figure}[h]
\begin{center}
\includegraphics[scale=0.20]{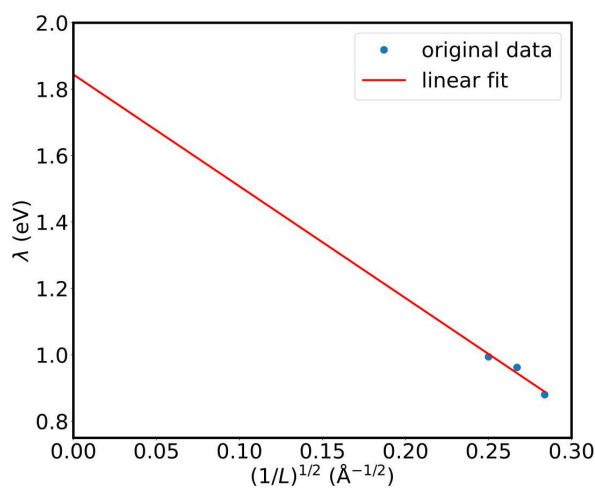}% Here is how to import EPS art
\caption{Linear fit of the reorganization free energy ($\lambda$) with $(1/L)^{1/2}$.
%Original data points are shown by solid circles and are fitted with a straight line (solid line).
%The error in the linear fit is indicated with the $R^2$ value.
}
\label{size_effect}
\end{center}
\end{figure}

In an alternative approach, one performs a series of simulations with increasing system size, 
%and calculate $\lambda$ for all these systems.
%
and $\lambda$ is computed for different system sizes, in particular, different simulation box lengths $L$.
%and plotted versus $1/L$.
%to make a linear fit of $\lambda$ with $(1/L)$.
%
 $\lambda$ is then plotted versus $1/L$ and
the extrapolated linear fit to $1/L \rightarrow 0$ gives the value of $\lambda$ at infinite dilution.
%and the $\lambda_{\text C}$ term can be calculated easily from this plot.
%
Blumberger and co-workers\cite{blumberger_Ru_ET_JACS} have shown that  such a linear fit with $1/L$ can underestimate the finite size correction.
As a better alternative, they proposed to fit using $(1/L)^{1/2}$.
}
%and obtained more accurate estimation for $\lambda_{\text C}$.}
%
%
%
%Following this strategy, 
To measure the finite size effects of our results using this approach, we carried out two additional simulations with bigger unit cells as discussed in the \nameref{methods_and_models} section.
In \cref{tab3}, we presented our results for these two systems with 92 and 137 water molecules at the level of BLYP functionals.
%
%We combined these results with the BLYP results of the 64 water system and obtained three data points for $\lambda$ with different cell sizes.
%
Together with the data from the 64 water simulation, 
%In order to find the correction for finite size, in \cref{size_effect}, 
we plot the reorganization free energy ($\lambda$) as a function of $(1/L)^{1/2}$ (\cref{size_effect}).
We used linear regression to fit these data points with a straight line.
The $R^2$ value of our fit is 0.936, which is reasonable, considering the few data points considered here.
%not as good as our earlier fits.
%
%This poor fitting is a result of having a limited number of data points.
%
By extrapolating 
%the line 
to the infinite dilution limit, we found 
%the correction term to be as Y eV.
%
that the value of $\lambda$ at infinite dilution is 1.84 eV.
%
%while the calculated value of $\lambda$ is 0.88 eV in our simulation with the smallest unit cell.
%
%The difference between these two $\lambda$ values gives the estimate of finite size correction ($\lambda_{\text C}$) in our calculation.
%
Thus, in the computed $\lambda$ values with 64 water molecules, we have added the correction term of 0.96~eV,
{The corrected results are in good agreement with the value of $\lambda_{\text C}$ = 0.91 eV computed with \cref{correction}.}
With this correction, our estimates of $\lambda$ are 1.84 eV and 2.06 eV for BLYP (GGA) and B3LYP (hybrid) based simulations.
The experimental value of $\lambda$ for this redox reaction is 2.1 eV.\cite{lambda_expt}
Thus, our hybrid functional based prediction of $\lambda$ is in excellent agreement with the experimental result.
%
%However, this apparent agreement should be considered with special care.
%
%since we have only three data points in \cref{size_effect} and the statistical errors of these data points were ignored.
%
%Moreover, we only considered 92 and 137 water molecules for our bigger systems, which might not be large enough to capture the solvent effect accurately.
%
%For a better estimate of the finite size correction, classical MD simulations with very large systems can be performed,\cite{blumberger_Ru_ET_JACS,sprik_M_ET_JPCB,size_effect_sprik}
%
%which is beyond the scope of our current work.
%
%Nevertheless, our results show that hybrid functional based AIMD simulations are inevitable for a better prediction of redox properties of aqueous electron transfer reactions.

Comparison of the computed redox potential $\Delta F$ with the experimental result is also not straightforward.
%
%In order to compare the results, 
First, finite size correction has to be applied  and
 %the difference in the definition of 
 the same experimentally considered zero electrostatic potential reference  has to be used in computations.\cite{blumberger_Ru_ET_JACS}
The finite size correction for $\Delta F$ is often evaluated by a linear fit of $\Delta F$ with $(1/L)^3$.\cite{blumberger_Ru_ET_JACS,sprik_M_ET_JPCB}
Then, the extrapolated linear fit can be used to obtain the value of $\Delta F$ at the infinite dilution limit.
However, our results suggest that the change in $\Delta F$ is minimal with the system size (see \cref{tab2,tab3}).
Thus, we refrain from computing the finite size correction for $\Delta F$.
Additionally, it is not trivial to compute the difference in the absolute electrostatic potential reference between our periodic DFT calculation and the experiment.
Hence, we could not compare the computed redox potential directly with the experimental results.
Regardless of the corrections, we observe that the value of $\Delta F$ for the BLYP case is 0.24 eV larger than for B3LYP, and that highlights the effect of the functional.

\section*{Conclusion}
\label{conclusion}
	
In this work, we studied the aqueous Fe$^{2+}$/Fe$^{3+}$ redox system employing the KS-DFT based AIMD technique.
Due to the SIE, GGA functionals are known to introduce errors in the 
redox properties of such open-shell systems.
Hybrid functional based AIMD simulations are better for such studies, as SIE is minimized because of the inclusion of some percentage of HF exchange.
However, hybrid functional based AIMD simulations with PW basis set are rarely performed due to the computational cost involved.
To overcome this, we performed NSMD based AIMD simulations wherein the HF exchange was computed using screened SCDM-based localized orbitals.
With this approach, we are able to generate long AIMD trajectories of aqueous Fe$^{2+}$/Fe$^{3+}$ redox systems containing $\sim$200 atoms.

Using our simulation results, we compared the accuracy of BLYP (GGA) and B3LYP (hybrid) functionals for studying this reaction.
%are benchmarked for the prediction of redox properties.
%
Although we observed negligible differences in the first solvation shell structure of the ions, the second solvation shell structure showed prominent differences.
%are more prominent.
%
To compute the redox properties, we followed the half-cell reaction modelling strategy by Sprik and co-workers employing AIMD simulations.
Specifically, we computed the redox potential and the solvent reorganization free energy ($\lambda$) using the vertical gap energy as the reaction coordinate to model the Fe$^{2+}$/Fe$^{3+}$ redox reaction.

We found that the free energy curves for the reduced and oxidized state of the system can be well approximated by two parabolas with the same curvature in accordance to the linear response assumption of Marcus theory.
%
%
%
%The comparison of the hybrid and GGA functional based results shows that 
The BLYP GGA functional predicted a 0.22~eV lower
solvent reorganization energy ($\lambda$)
than the B3LYP hybrid functional.
However, both functionals underestimate the solvent reorganization energy by $\sim$ 1~eV as compared to the experimental observations.
%
%this discrepancy in the predicted $\lambda$ is due to the finite size of our simulation cell, which is unable to take into account the solvent effect completely.
%
%Howevr, we noticed that the correction fpr the finite size is large.
%
Following the earlier reports, we computed the finite size corrections.
%we performed two additional simulations with larger unit cells at the level of the BLYP (GGA) functional and found the correction term to be about 0.96~eV.
%
After including this correction, amounting to 0.96~eV, the computed value of $\lambda$ at the level of the hybrid functionals agrees well with the experimental data.
These results demonstrate the importance of using hybrid functional based AIMD for the accurate prediction of redox properties in open-shell systems.

The work also shows that the NSMD technique along with screened localized orbitals is a good approach for speeding up hybrid density functional based AIMD simulations of complex chemical reactions in water. 
The protocols used here can also be used for studying other electron transfer reactions.

%that the computationally demanding AIMD simulations with hybrid functional are only practically possible because of the speed up achieved by the NSMD method.
%
%
%Also, we studied the effect of solvent size for the prediction of reorganization energies.
%
%Nevertheless, in spite of these limits, our observation suggest the superiority of the hybrid functional for these kind of studies.

%\clearpage

\section*{Acknowledgements}
The support and the resources provided by the Centre for Development of Advanced Computing (C-DAC), the National Supercomputing Mission (NSM), Government of India, the  Science and Engineering Research Board (India) under the MATRICS (Ref. No. MTR/2019/000359), the German Research Foundation (DFG) through SFB 953 (project number 182849149), the Federal Ministry of Education and Research (BMBF), the state of Bavaria as part of the NHR Program, the Cluster of
Excellence "Engineering of Advanced Materials" (EAM)
and the "Competence Unit for Scientific Computing"
(CSC) at the University of Erlangen-N{\"u}rnberg (FAU)
are gratefully acknowledged.
RK thanks the Council of Scientific \& Industrial Research (CSIR), India, for her PhD fellowship.
Computational resources were provided by the HPC facility (HPC2013)
at IITK, the Erlangen Regional Computing Center (RRZE) at FAU,
SuperMUC-NG (project pn98fa) at Leibniz Supercomputing Centre (LRZ)
and PARAM Sanganak supercomputing facility under NSM at IITK.

\section*{Conflict of Interest}

The authors declare no conflict of interest.

%%%%%%%%%%%%%%%%%%%%%%%%%%%%%%%%%%%%%%%%%%%%%%%%%%%%%%%%%%
%%%%%%%%%%%%%%%%%%%%%%%%%%%%%%%%%%%%%%%%%%%%%%%%%%%%%%%%%%
%%%%%%%%%%%%%%%%%%%%%%%%%%%%%%%%%%%%%%%%%%%%%%%%%%%%%%%%%%
\begin{shaded}
\noindent\textsf{\textbf{Keywords:} \keywords} 
\end{shaded}
%%%%%%%%%%%%%%%%%%%%%%%%%%%%%%%%%%%%%%%%%%%%%%%%%%%%%%%%%%
%%%%%%%%%%%%%%%%%%%%%%%%%%%%%%%%%%%%%%%%%%%%%%%%%%%%%%%%%%
%%%%%%%%%%%%%%%%%%%%%%%%%%%%%%%%%%%%%%%%%%%%%%%%%%%%%%%%%%

%%%%%%%		References			%%%%%%% 

\setlength{\bibsep}{0.0cm}
\bibliographystyle{Wiley-chemistry}
\bibliography{example_refs}

%%%%%%%		TOC Entry			%%%%%%% 

%%%%%%%%%%%%%%%%%%%%%%%%%%%%%%%%%%%%%%%%%%%%%%%%%%%%%%%%%%
%%%%%%%%%%%%%%%%%%%%%%%%%%%%%%%%%%%%%%%%%%%%%%%%%%%%%%%%%%
%%%%%%%%%%%%%%%%%%%%%%%%%%%%%%%%%%%%%%%%%%%%%%%%%%%%%%%%%%

\end{document}